%% file: main.tex
\documentclass[conference]{IEEEtran}
\IEEEoverridecommandlockouts
\usepackage{cite}
\usepackage{amsmath,amssymb,amsfonts}
\usepackage{algorithmic}
\usepackage{graphicx}
\usepackage{textcomp}
\usepackage{xcolor}

\usepackage[linesnumbered,ruled]{algorithm2e}
\def\BibTeX{{\rm B\kern-.05em{\sc i\kern-.025em b}\kern-.08em
    T\kern-.1667em\lower.7ex\hbox{E}\kern-.125emX}}
\begin{document}

\title{
\huge{A-DRIVE: Autonomous Deadlock Detection and Recovery\\at Road Intersections for Connected and Automated Vehicles}}

\author{\IEEEauthorblockN{Shunsuke Aoki\IEEEauthorrefmark{1},
Ragunathan (Raj) Rajkumar\IEEEauthorrefmark{2}
}
\IEEEauthorblockA{\IEEEauthorrefmark{1}National Institute of Informatics, Tokyo, Japan}
\IEEEauthorblockA{\IEEEauthorrefmark{2}Carnegie Mellon University, Pittsburgh, PA USA}
}

\maketitle

\begin{abstract}
\input{00_Abstract}%OK -0702

\end{abstract}


\section{Introduction}

\input{01_Introduction}%OK -0702

\section{Assumptions}

\input{02_Assumptions}

\section{A-DRIVE: A Deadlock Detection and Recovery Protocol}
\input{03_Protocol}

\vspace{3mm}

\section{Implementation and Evaluation}
\input{04_Implementation}

\vspace{3mm}

\section{Related Work}
\input{05_relatedwork}

\vspace{3mm}
\section{Conclusion}
\input{06_conclusion}

\bibliographystyle{./bibliography/IEEEtran}
\bibliography{main}

\end{document}

%% file: 00_Abstract.tex
Connected and Automated Vehicles (CAVs) are highly expected to improve traffic throughput and safety at road intersections, single-track lanes, and construction zones.
However, multiple CAVs can block each other and create a mutual deadlock around these road segments (i) when vehicle systems have a failure, such as a communication failure, control failure, or localization failure and/or (ii) when vehicles use a long shared road segment.
In this paper, we present an Autonomous Deadlock Detection and Recovery Protocol at Intersections for Automated Vehicles named \textit{A-DRIVE} that is a decentralized and time-sensitive technique to improve traffic throughput and shorten worst-case recovery time.
To enable the deadlock recovery with automated vehicles and with human-driven vehicles, A-DRIVE includes two components: \textit{V2V communication-based A-DRIVE} and \textit{Local perception-based A-DRIVE}.
V2V communication-based A-DRIVE is designed for homogeneous traffic environments in which all the vehicles are connected and automated.
Local perception-based A-DRIVE is for mixed traffic, where CAVs, non-connected automated vehicles, and human-driven vehicles co-exist and cooperate with one another.
Since these two components are not exclusive, CAVs inclusively and seamlessly use them in practice.
Finally, our simulation results show that A-DRIVE improves traffic throughput compared to a baseline protocol.

%% file: 01_Introduction.tex
Vehicle cooperation is becoming more important for Connected and Automated Vehicles (CAVs) to improve road safety and traffic throughput.
In fact, according to the National Highway Traffic Safety Administration (NHTSA), more than $35,000$ people die in motor vehicle-related crashes in the US every year \cite{NHTSA-AutomatedSafety}.
Automation technologies can reduce that number because more than $90 \%$ of serious crashes occur due to human error \cite{NHTSA-AutomatedSafety}.
Hence, there are a variety of applications to improve road safety with vehicular cooperation and automation \cite{cecchini2017performance, wang2017cooperative, aoki2021comm_Magazine}.

Safe cooperation and collaboration among multiple vehicles is one of the prospective applications for road safety, when conflicts arise on shared road segments, such as \textit{stationary intersections} \cite{azimi2014stip, lin2019graph, aoki2017configurable, gunther2019optimizing} and \textit{dynamic intersections} \cite{aoki2018dynamic}, including merge points, single-track lanes, and construction zones \cite{zhao2019state, aoki2017merging}.
Under these cooperative protocols, each CAV uses Vehicle-to-Infrastructure (V2I) Communications and/or Vehicle-to-Vehicle (V2V) Communications for inter-vehicle interaction and negotiation to avoid vehicle collisions and deadlocks. Also, each CAV uses on-board perception systems to guarantee road safety.

\begin{figure}[!t]
  \begin{center}
    \begin{tabular}{c}
      % 1
      \begin{minipage}{0.5\hsize}
        \begin{center}
          \includegraphics[clip, width=3.75cm]{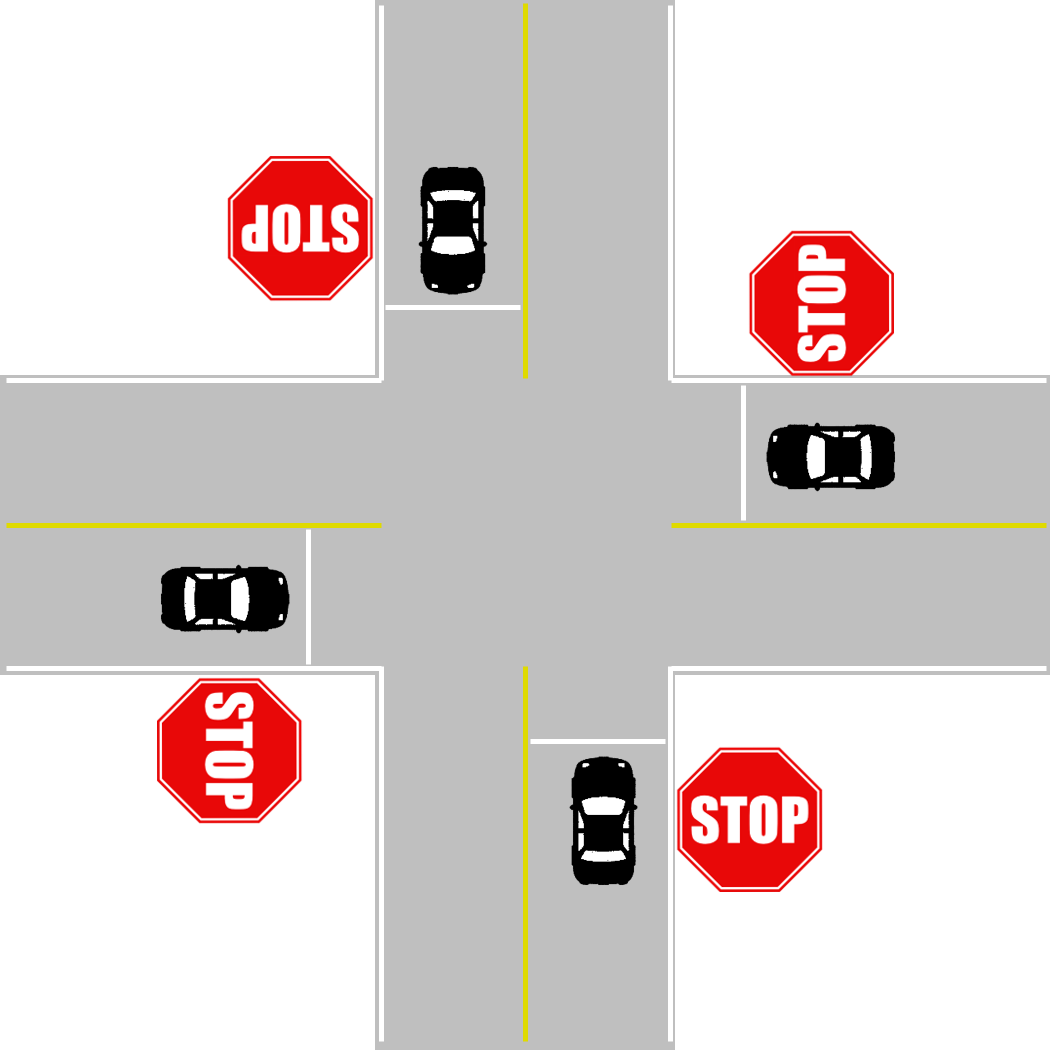}
	\hspace{1.6cm} (a) Priority Miscalculation\\at 4-way Intersection.
        \label{fig:case1}
        \end{center}
      \end{minipage}
      % 2
      \begin{minipage}{0.5\hsize}
        \begin{center}
          \includegraphics[clip, width=3.75cm]{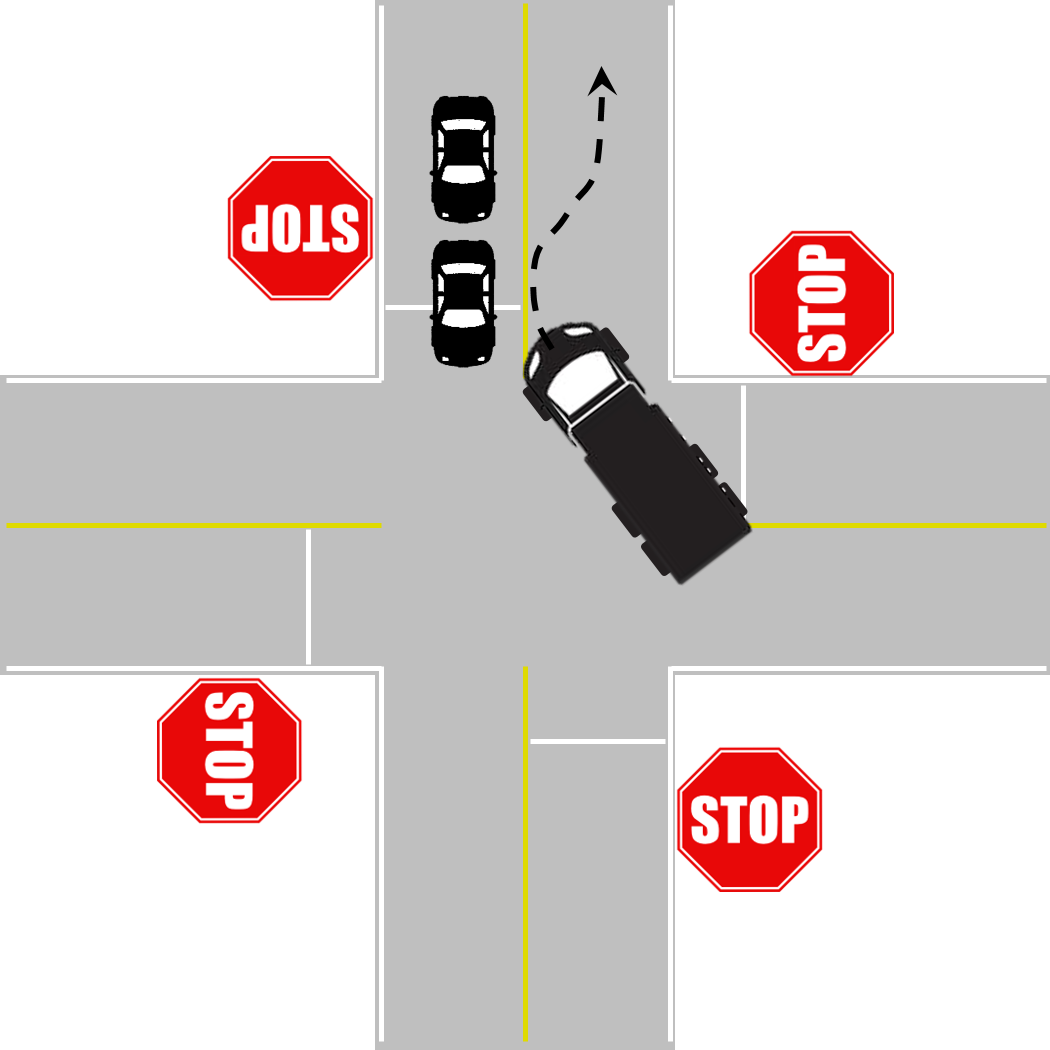}
	\hspace{1.6cm} (b) Exceeding Stop Line\\at 4-way Intersection.
          \label{fig:case2}
        \end{center}
      \end{minipage}\\
      \begin{minipage}{0.10\hsize}
      \end{minipage}\\
      % 3
      \begin{minipage}{0.5\hsize}
        \begin{center}
          \includegraphics[clip, width=3.75cm]{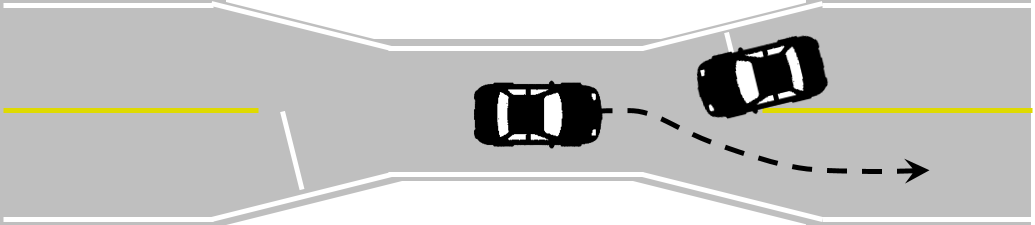}
	\hspace{1.6cm} (c) Exceeding Stop Line\\at Narrow Road.
          \label{fig:case2}
        \end{center}
      \end{minipage}
      % 4
      \begin{minipage}{0.5\hsize}
        \begin{center}
          \includegraphics[clip, width=2.75cm]{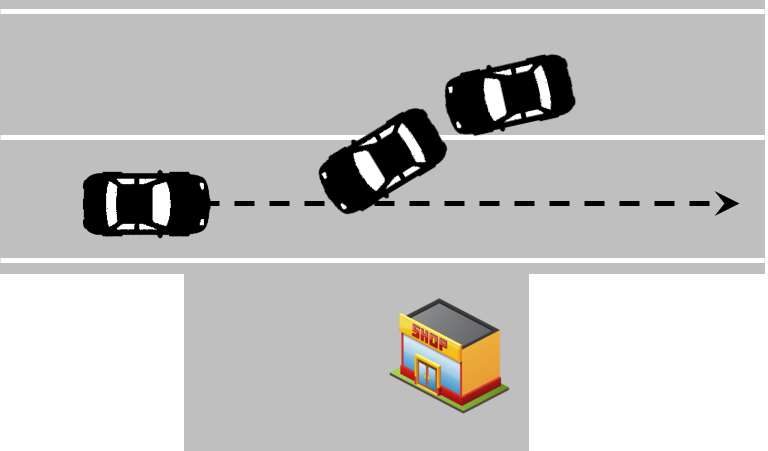}
	\hspace{1.6cm} (d) Exceeding Virtual Stop Line before Turning Left.\\
          \label{fig:case2}
        \end{center}
      \end{minipage}\\
      \begin{minipage}{0.1\hsize}
      \end{minipage}\\
      % 5
      \begin{minipage}{1.0\hsize}
        \begin{center}
          \includegraphics[clip, width=7cm]{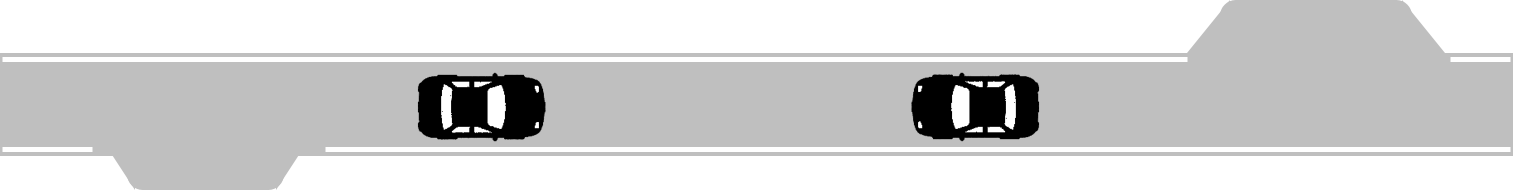}
	\hspace{1.6cm} (e) Long Single-track Lane.
          \label{fig:case2}
        \end{center}
      \end{minipage}

    \end{tabular}
    \caption{Deadlock on Public Roads.}
\label{fig:Scenarios4Deadlock}
  \end{center}
\end{figure}

However, even when CAVs follow these cooperative protocols, they might create mutual deadlock \cite{jager2001decentralized, holt1972some}, (i) when vehicle systems have a failure, such as a control failure, communication failure, or localization failure and/or (ii) when vehicles traverse a long \textit{critical section}.
We present $5$ example scenarios for vehicle deadlocks in Figure \ref{fig:Scenarios4Deadlock}.
For example, as shown in Figure \ref{fig:Scenarios4Deadlock}-(a), the vehicles arrive at the intersection nearly at the same time, and each vehicle considers that it is the lower-priority vehicle in the FCFS (First-Come First-Served) -based coordination, which may lead to a deadlock. This may happen due to the localization error and/or time-synchronization error. To break a tie, at least one vehicle has to take an action and proceed.
Also, as shown in Figures \ref{fig:Scenarios4Deadlock}-(b), -(c) and -(d), when one vehicle goes beyond the stop line due to a control failure or a localization failure, the vehicle keeps waiting and blocking the on-coming traffic, which causes the deadlock situation.
In particular, in Figure \ref{fig:Scenarios4Deadlock}-(d), there are no explicit stop line in the map database but each vehicle calculates and estimates the virtual stop line. When the on-coming traffic includes a larger vehicle, the waiting vehicle has to re-calculate and recede to make a sufficient space.
These 4 examples do not happen with the perfect systems that have no control, communication, localization, and/or time-synchronization failures, but the practical vehicle systems can lead these deadlock situations.
In Figure \ref{fig:Scenarios4Deadlock}-(e), the vehicles can detect each other after entering the critical section, because the area is too long to entirely cover by local perception systems and/or vehicular communication systems.
One of them has to yield and/or enter an evacuation space. Here, such evacuation space for two-way traffic at single-track lanes are generally called \textit{passing places} or \textit{turnouts} \cite{HighwayCodeUK}.

In this paper, to solve these deadlock problems, we present an Autonomous Deadlock Detection and Recovery Protocol at Intersections for Automated Vehicles named \textit{A-DRIVE} that is a decentralized and time-sensitive technique to improve traffic throughput and shorten worst-case recovery time.
In A-DRIVE, each CAV uses V2V communications and on-board perception systems and determines which vehicle yields and/or recedes.
A-DRIVE includes two components: \textit{V2V communication-based A-DRIVE} and \textit{Local perception-based A-DRIVE}.
V2V communication-based A-DRIVE is designed for homogeneous traffic environments in which all the vehicles are connected and automated.
In this component, all vehicles use V2V communications to negotiate with each other.
Local perception-based A-DRIVE is developed for mixed traffic, where CAVs, non-connected automated vehicles, and human-driven vehicles co-exist and cooperate with one another.
Since there might be a long transition period before all human-driven vehicles can be replaced with automated vehicles \cite{bansal2017forecasting}, automated vehicles have to co-exist with human-driven vehicles and have to safely navigate themselves around each other \cite{aoki2021comm_Magazine}.
These $2$ components are not exclusive and each CAV inclusively and seamlessly uses them in practice.
In fact, in these $2$ components, each CAV uses common vehicle state, localization state, and cost function, to determine which vehicle has the priority to use the shared road segment where the deadlock happens.
When all the vehicles involved in a deadlock situation are confirmed as CAVs, each CAV uses the V2V communication-based A-DRIVE.
When a non-connected vehicle is detected, each CAV uses the local perception-based protocol.
The priority assignment schemes are common for these $2$ components.

The key contributions of this paper are as follows:

\begin{itemize}
\item We analyze and classify deadlock situations that worsen traffic throughput at stationary and dynamic intersections.
\item We present an Autonomous Deadlock Detection and Recovery Protocol at Intersections for Automated Vehicles named \textit{A-DRIVE}, where CAVs is able to recover from the deadlock situations in a decentralized manner.
\item We evaluate A-DRIVE using a simulator-emulator and show that our protocol delivers better performance in terms of throughput and recovery time.
\end{itemize}

%% file: 02_Assumptions.tex
In this section, we present a classification of vehicle deadlocks and review two different types of traffic environments to consider the interaction and cooperation between human-driven vehicles and automated vehicles.

\subsection{Vehicle Deadlocks}

A vehicle deadlock is a state where each vehicle is waiting for another vehicle around and/or inside shared road segments, such as four-way stationary intersections, merging points, single-track lanes, and construction zones.
To recover from the vehicle deadlock, vehicles have to break a tie and/or at least one vehicle has to yield and recede.

The vehicle deadlock happens even when the centralized and decentralized cooperation protocols \cite{azimi2014stip, lin2019graph, aoki2017configurable, aoki2018dynamic} work properly, due to two reasons: (i) System Failure and (ii) Large Intersection.

\textit{\textbf{Deadlock due to System Failure:}} This deadlock happens due to a control failure, a communication failure, a localization failure, and/or a time-synchronization failure of the vehicle system. Under this deadlock, at least one vehicle fails to stop at the appropriate location and may violate the critical section.

\textit{\textbf{Deadlock due to Large Intersection:}} At a large intersection, such as a long single-track lane and/or large center turn lane, each vehicle cannot detect and/or negotiate with each other until deadlock happens because of the limitations of communication ranges and/or sensor-detection ranges.

These deadlocks happen even for human drivers.
Human drivers recover from these deadlocks by accounting for the surrounding situations.
For example, when one vehicle exceeds the stop line, as shown in Figures \ref{fig:Scenarios4Deadlock}-(b), -(c), and -(d), the vehicle has to yield and recede to allow the on-coming traffic to move forward. Also, for the case shown in Figure \ref{fig:Scenarios4Deadlock}-(e), the usage of single-track lanes and passing places are explicitly described in the Highway Code Book at UK \cite{HighwayCodeUK}.
To guarantee road safety, the drivers use a combination of traffic rules, social norms, and verbal/non-verbal communications.

\subsection{Traffic Environments}

Our deadlock recovery protocol A-DRIVE is designed for two traffic environments, \textit{Homogeneous Traffic} and \textit{Mixed Traffic}, keeping in mind that a long transition period might be required \cite{bansal2017forecasting}.

\textit{\textbf{Homogeneous Traffic:}} The traffic only consists of Connected and Automated Vehicles (CAVs). In our protocol, V2V communication-based A-DRIVE is designed for this traffic environment.

\textit{\textbf{Mixed Traffic:}} The traffic has CAVs as well as non-connected self-driving vehicles and human-driven vehicles. These non-connected vehicles are not equipped with any communication devices.
The local perception-based A-DRIVE component is for this traffic environment.

Even after the spread of autonomous driving technologies in the society, human-driven vehicles might remain for a while. Therefore, we have to design a vehicle cooperation protocol both for the homogeneous traffic and for the mixed traffic, and these protocols should be seamlessly used.

%% file: 03_Protocol.tex
In this section, we present our cooperative deadlock detection and recovery protocol named A-DRIVE.
The protocol is designed to solve and recover from vehicle deadlocks by using Vehicle-to-Vehicle (V2V) communications and on-board perception systems.
Since the vehicle deadlocks can happen almost anywhere and anytime on public roads, we use V2V communications, local on-board perception, and decentralized mechanisms.
A-DRIVE includes $2$ subcomponents: V2V communication-based A-DRIVE and Local perception-based A-DRIVE. They can be used in a mixed traffic environment and in a homogeneous traffic environment seamlessly.
Specifically, the decision-making processes in these $2$ components are same and they use common state transitions and cost function.
Each CAV calculates its threshold waiting time in a deadlock situation based on its state and yielding cost in a decentralized manner.
Once its waiting time exceeds the given threshold, the vehicle starts the action to resolve the deadlock.

Only the differences between these $2$ components are negotiation processes.
In V2V communication-based A-DRIVE, each CAV exchanges the message over V2V communications and negotiates the priority to use the shared road segment.
On the other hand, in Local perception-based A-DRIVE, each CAV immediately starts the backing operation once its waiting time exceeds the given threshold time.
The local perception systems always check the movements of other vehicles, and therefore, the lower-priority vehicles start the yielding maneuvers first and the higher-priority vehicle might be remained and consequently the higher-priority vehicle can use the shared road segment first.

In addition, we use the yielding cost function and cost-based priority assignment, in order to account for the physical constraints and traffic.
A simple exponential backoff used in Ethernet and computer networks \cite{7428776} is neither efficient nor optimal because of the background features.
For example, the vehicle movements to yield and recede are time-consuming maneuvers, unlike the backoff in computer networks.
Also, when one vehicle blocks any on-coming multiple vehicles and a deadlock happens, the vehicle has to yield and recede, because the backing operation of multiple vehicles is generally difficult and time- and fuel-consuming.

From a standards viewpoint, we use the second optional part of Basic Safety Messages (BSM) \cite{kenney2011dedicated} for Advanced Safety Messages.
Therefore, Advanced Safety Messages are broadcast at 10 Hz, the same rate as the BSM.
BSM includes the vehicle basic information, such as position coordinates, heading and velocity, and has the second part that is configurable for application providers.

Since CAVs are safety-critical and life-critical applications, our cooperative deadlock recovery protocol includes safe mechanisms and never compromise road safety.

\begin{figure}[!t]
\centering
\includegraphics[width=6.0cm]{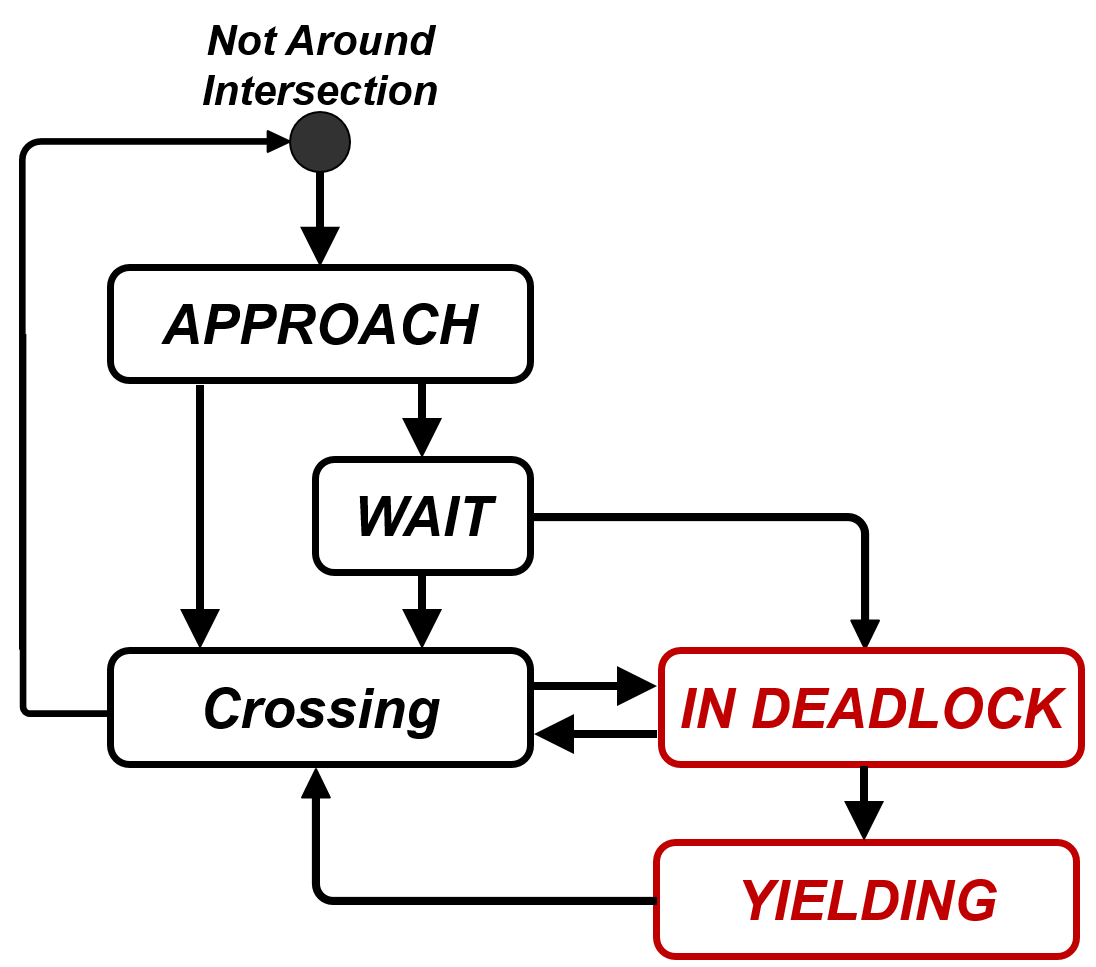}
\caption{Vehicle State Transition used by Vehicles at Intersections.}
\label{fig:state_deadlockrecovery}
\end{figure}

\subsection{State Transitions and Cost Function}\label{sec:yieliding_cost}

%Done: July 30
In this section, we present a priority assignment technique that determines which vehicle can proceed and which vehicle must yield and recede to resolve vehicle deadlocks.
This component is commonly used for V2V communication-based A-DRIVE and for Local perception-based A-DRIVE. In these $2$ A-DRIVE approaches, we use $3$ elements: (i) vehicle state, (ii) localization state, and (iii) yielding cost.

\begin{figure}[!b]
  \begin{center}
    \begin{tabular}{c}
      % 1
      \begin{minipage}{0.5\hsize}
        \begin{center}
          \includegraphics[clip, width=3.85cm]{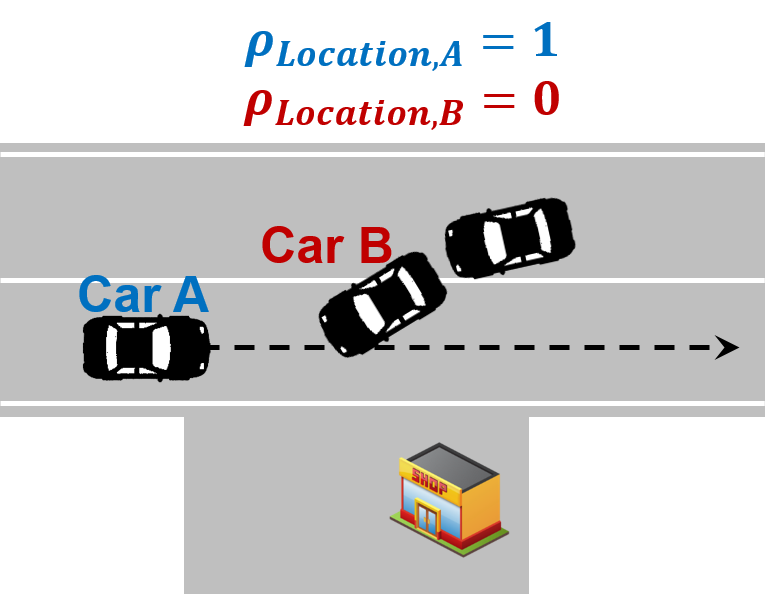}
	\hspace{1.6cm} (a) Vehicle Exceeding\\Virtual Stop Line.
        \label{fig:case1}
        \end{center}
      \end{minipage}
      % 2
      \begin{minipage}{0.5\hsize}
        \begin{center}
          \includegraphics[clip, width=3.45cm]{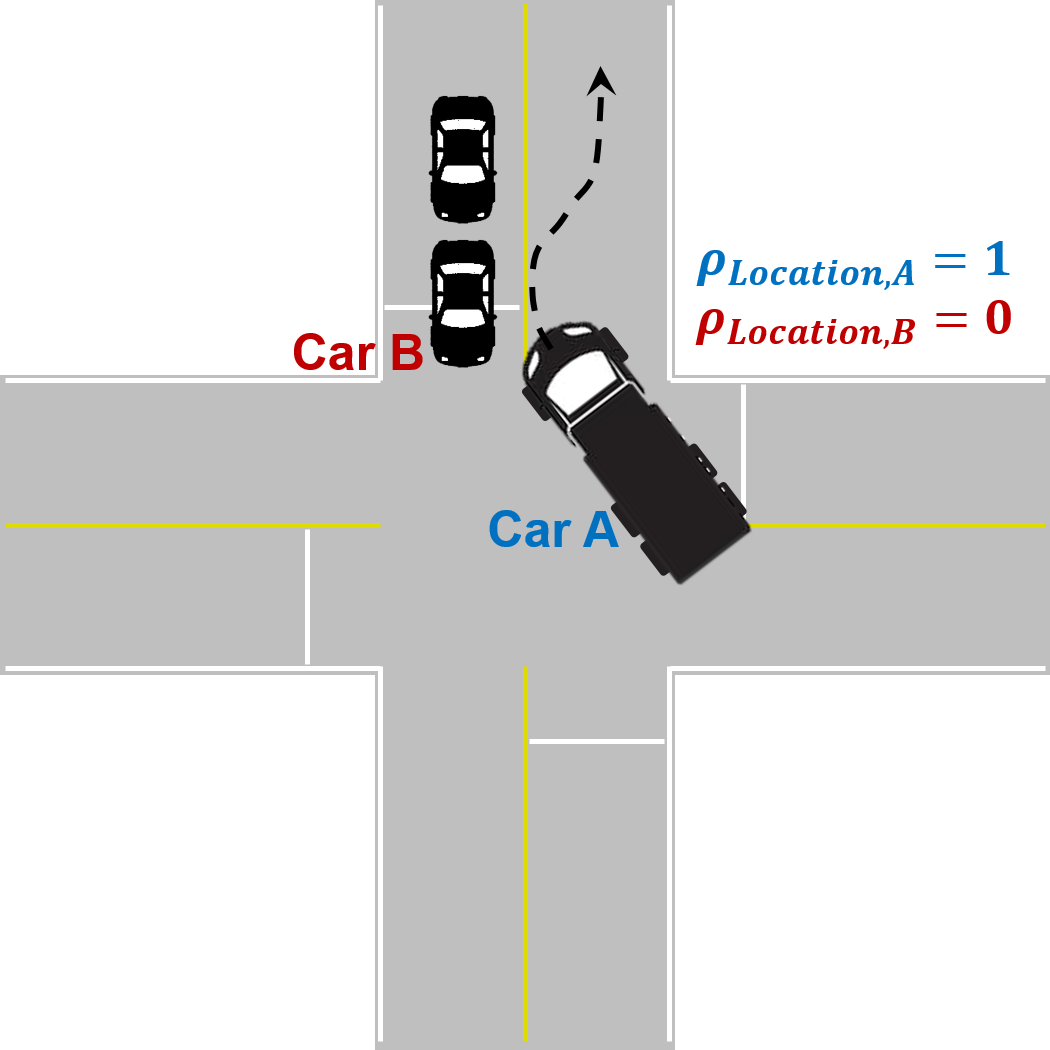}
	\hspace{1.6cm} (b) Vehicle Exceeding\\Stop Line.
          \label{fig:case2}
        \end{center}
      \end{minipage}\\
      \begin{minipage}{0.10\hsize}
      \end{minipage}\\
    \end{tabular}
    \caption{Localization State and Vehicle Deadlocks.}
\label{fig:LocalizationState}
  \end{center}
\end{figure}

First, each vehicle is in a vehicle state $\Phi_{state}$ that dynamically transitions, and is used to make decisions for vehicle navigation.
Secondly, each vehicle holds its \textit{localization state} $\rho_{location}$ that represents the location recognized by the vehicle. The localization state is a Boolean function and becomes $1$ when the vehicle considers its location is within the intersection or within the shared road segment. On the other hand, the localization state $\rho_{location}$ becomes $0$ if the vehicle is outside the intersection area.
This value is very useful to resolve a vehicle deadlock, because the deadlock may happen due to the miscalculation of the vehicle location and/or vehicle state.
For example, in Figures \ref{fig:LocalizationState}-(a) and -(b), Car B waits at the stop line but exceeds the given line due to the localization errors and/or control failures. Car B are actually within the intersection but it considers its location is out of the intersection area and its localization state $\rho_{location}$ becomes $0$.
Thirdly, each vehicle calculates its \textit{yielding cost} $\chi(v_i)$ that represents its cost to yield for resolving the vehicle deadlock.
The cost $\chi(v_i)$ becomes larger value when the yielding maneuver for vehicle $i$ is more time-consuming.

Each vehicle uses these $3$ elements to negotiate with surrounding vehicles and to resolve the vehicle deadlock.
We first present these $3$ elements and secondly discuss the priority assignment and tie-breaking rules.

\subsubsection{Vehicle States and Localization States}\label{sec:state}

%Done: July 30
For deadlock recovery protocols, we use $6$ vehicle states: \textit{NOT AROUND INTERSECTION}, \textit{APPROACH}, \textit{WAIT}, \textit{CROSSING}, \textit{IN DEADLOCK} and \textit{YIELDING}.
The state transition diagram is shown in Figure \ref{fig:state_deadlockrecovery}.
We briefly describe these vehicle states here.

\begin{description}
 \item[1. \textit{NOT AROUND INTERSECTION}]\mbox{}\\
Any vehicles not near the stationary and/or dynamic intersections are in the \textit{NOT AROUND INTERSECTION} state.

 \item[2. \textit{APPROACH}]\mbox{}\\ 
When the vehicle approaches the intersection, its state $\Phi_{state}$ transitions to the \textit{APPROACH} state.

 \item[3. \textit{WAIT}]\mbox{}\\ 
After the vehicle reaches and stops at the stop line or virtual stop line at the intersection, its state becomes \textit{WAIT} state. The vehicle in the \textit{WAIT} state is not inside the stationary and/or dynamic intersection when the system has no localization failure.

 \item[4. \textit{CROSSING}]\mbox{}\\ 
Once the vehicle enters the stationary and/or dynamic intersection, the vehicle state $\Phi_{state}$ transitions to the \textit{CROSSING} state.
After the vehicle exits the intersections, its state transitions to the \textit{NOT AROUND INTERSECTION} state.

\item[5. \textit{IN DEADLOCK}]\mbox{}\\ 
Once the vehicle detects that it is involved in a vehicle deadlock, the vehicle state transitions to the \textit{IN DEADLOCK} state.
Each vehicle can detect the deadlock by using its local perception and its wait time. In particular, the vehicles in the \textit{WAIT} state and the \textit{CROSSING} state have to keep sensing to detect the vehicle deadlock.

\item[6. \textit{YIELDING}]\mbox{}\\ 
When the vehicle uses the deadlock recovery protocol and determines to yield and recede, the vehicle state transitions to the \textit{YIELDING} state.

\end{description}

Each CAV holds $1$ of these $6$ vehicle states to negotiate with the other vehicles. In practice, vehicles in the \textit{WAIT} and \textit{CROSSING} can be involved in the vehicle deadlocks.

\subsubsection{Follower-aware Cost Function}\label{sec:costfunc}

%Done: July 31
In this section, we present a \textit{follower-aware cost function} that represents the cost to yield to resolve a vehicle deadlock.
The follower-aware cost function considers $2$ factors: (i) Distance to a passing space or to the entrance of the intersection and (ii) Presence of the following vehicles.

%Done: July 31
First, as shown in Figure \ref{fig:Scenarios4Deadlock}-(e), when two vehicles face each other in a critical section, the vehicle closer to a passing space should yield and recede. If there are no passing places or wide parts of the road, the entrance of the critical section is considered as the evacuation site.
This approach contributes to shorten the time to resolve the deadlock.

%Done: July 31
Secondly, when one vehicle blocks on-coming multiple vehicles and leads a deadlock, it is preferable that the single vehicle yields and recedes, because the backing operation of multiple vehicles might be difficult and time-consuming.

%Done: July 31
As human drivers comprehensively understand the situation and resolve a vehicle deadlock, the follower-aware cost function gives the yielding cost to increase the traffic throughput.

%Done: July 31
We present the cost function and cost to yield $\chi$ in Eqs. (\ref{eq:cost_func_comm}) and (\ref{eq:cost_func_percp}).
Here, $\chi(v_{i, comm})$ and $\chi(v_{i, perception})$ represent the costs to yield for vehicle $i$ for V2V communication-based A-DRIVE and Perception-based A-DRIVE, respectively.

\begin{figure}[!t]
\centering
\includegraphics[width=7.35cm]{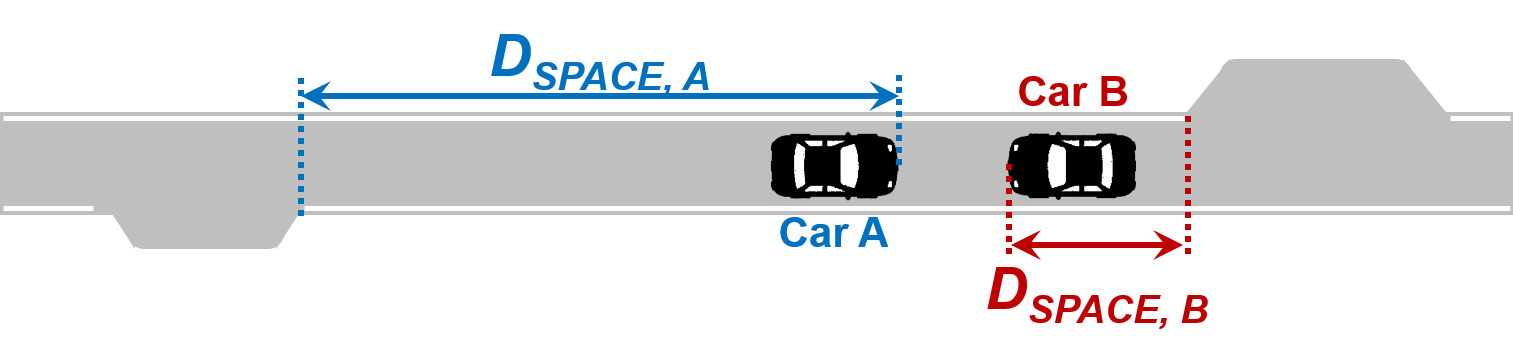}
\caption{Distance to a Passing Place and a Vehicle Deadlock.}
\label{fig:Dspaceexplain}
\end{figure}

\begin{figure}[!t]
\centering
\includegraphics[width=7.35cm]{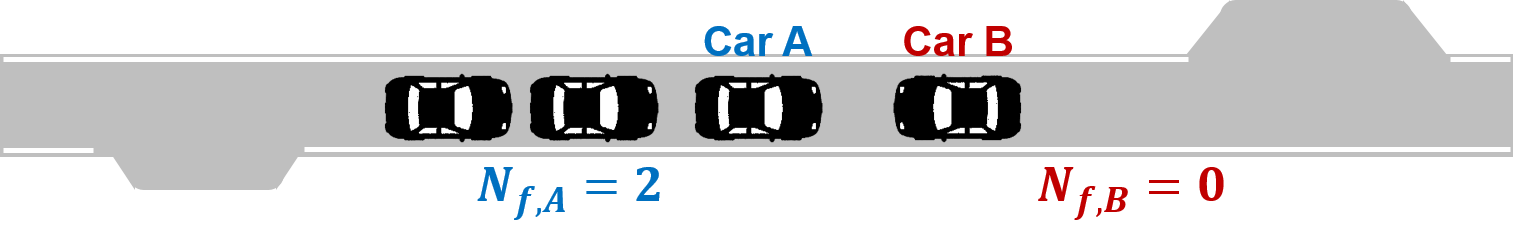}
\caption{Number of the Following Vehicles.}
\label{fig:Nf_number}
\end{figure}

\begin{equation}
\chi(v_{i, comm}) = s_{comm} \cdot D_{SPACE} + t_{comm} \cdot N_{f}
\label{eq:cost_func_comm}
\end{equation}

\begin{equation}
\chi(v_{i, perception}) = s_{perception} \cdot D_{SPACE} + t_{perception} \cdot G
\label{eq:cost_func_percp}
\end{equation}

In these equations, $D_{SPACE}$ describes the physical distance to a passing place and/or evacuation site, as shown in Figure \ref{fig:Dspaceexplain}. When the physical distance $D_{SPACE}$ is larger, the cost $\chi(v_i)$ becomes larger value.
For example, in Figure\ref{fig:Dspaceexplain}, Car A has larger value for the distance $D_{SPACE}$ and the yielding cost $\chi(v_i)$ becomes larger.
$N_f$ describes the number of the following vehicles within the communication range. Since CAV gets the information from the received safety beacon, BSM, they can get this value.
As shown in Figure \ref{fig:Nf_number}, Car A has two followers and $N_{f,A}$ becomes $2$. In addition, Car B has no followers and $N_{f,B}$ becomes $0$.
In fact, when we have a larger value for $N_f$, the yielding and backing operation become more difficult and more time-consuming.
$G$ is a Boolean function to describe the presence of the following vehicle, because each CAV cannot get the exact number of its followers only by using on-board perception systems.
Lastly, $s_{comm}$, $s_{perception}$, $t_{comm}$ and $t_{perception}$ are scale factors.

\subsubsection{Priority for Deadlock Recovery}

%Done: July 31
The protocol uses these $3$ factors, vehicle state $\Phi_{state}$, localization state $\rho_{location}$, and yielding cost $\chi$, to determine the priority and which vehicle yields and recedes.
We present $3$ rules of deadlock recovery for Connected and Automated Vehicles here.

\begin{enumerate}
 \item \textbf{Vehicles within the intersection ($\rho_{location}=1$) have the higher priority, compared to vehicles outside the intersection ($\rho_{location}=0$).}
 \item \textbf{When multiple vehicles block each other and they are within the intersection ($\rho_{location}=1$), the vehicle having the larger yielding cost has the higher priority.}
 \item \textbf{When multiple vehicles block each other and they are outside the intersection ($\rho_{location}=0$), the vehicles use the random numbers to break a tie.}
\end{enumerate}

%Done: July 31
We give these rules to guarantee road safety while increasing traffic throughput when CAVs resolve the deadlock.
First, the localization state is the most important factor. The vehicles within the intersection have the higher priority, because at least one vehicle has to proceed and exit the intersection in order to resolve the deadlock. The deadlock shown in Figures \ref{fig:Scenarios4Deadlock}-(b), -(c) and -(d), can be resolved by using this rule.
Secondly, since the yielding cost $\chi(v_i)$ represents the difficulty and complexity of backing operation of vehicle $i$, we can use the cost to determine the priority. The deadlock shown in Figure \ref{fig:Scenarios4Deadlock}-(e) can be resolved by using the yielding cost.
Thirdly, as shown in \ref{fig:Scenarios4Deadlock}-(a), when multiple vehicles wait for each other before the stop line at the intersection, they simply use the random numbers to resolve the deadlock. Overall, we use the vehicle state to detect the vehicle deadlock, and we mainly use the localization state and the yielding cost for the negotiation.

\subsection{V2V Communications-based A-DRIVE: For Homogeneous Traffic}

Here, we present V2V communication-based A-DRIVE that uses V2V communications for safe cooperation.

\subsubsection{Message Contents for V2V Communications}

To enable a peer-to-peer negotiation by using V2V communications, CAVs exchange $5$ elements as follows.

\begin{itemize}
  \item vehicle state $\Phi_{state}$.
  \item localization state $\rho_{location}$.
  \item yielding cost $\chi$.
  \item random number $R$.
  \item presence of human-driven vehicle(s) $HV_{flag}$
\end{itemize}

These elements are used for determining a priority and resolving a deadlock.

$HV_{flag}$ is the flag representing the presence of human-driven vehicle(s) in the vehicle deadlock.
When there are no human-driven vehicles involved in the deadlock, the flag becomes $0$ and each CAV follows V2V communication-based A-DRIVE.
On the other hand, when the flag becomes $1$, each CAV follows Local perception-based A-DRIVE.
To keep the consistency among multiple vehicles involved in the deadlock, when one vehicle has $1$ for the $HV_{flag}$, all the neighboring vehicles change its value to $1$.
In other words, CAVs cooperatively sense the presence of non-connected and human-driven vehicles to improve road safety.

\subsubsection{Detection of Deadlocks}

By definition, a vehicle deadlock happens when multiple vehicles wait for each other.
A CAV can detect the presence of a deadlock using its map database that shows other vehicle paths and using on-board perceptions and the received messages over V2V communications that show other vehicle states and locations.
Our deadlock recovery protocol will then be initiated and each vehicle negotiates to resolve the deadlock.

\subsubsection{Negotiation and Deadlock Recovery}\label{sec:comm_nego}

In the protocol, each CAV determines its priority by exchanging their vehicle states $\Phi_{state}$, localization states $\rho_{location}$, yielding costs $\chi$ and random value $R$.
First, the vehicle states are used for understanding which vehicles are involved in the vehicle deadlock.
Secondly, the localization states and yielding costs are used for the priority assignment.

When all vehicles have the same localization states, each vehicle has to negotiate by using the yielding costs. In this negotiation, the vehicle having the largest value on the yielding cost have the priority to use the intersection.
On the other hand, When vehicles having $0$ and $1$ in the localization state $\rho_{location}$ are mixed, the vehicles having $0$ in the state $\rho_{location}$ have to start the backing operation, because they are blocking other vehicles at the intersection.

In this V2V communication-based A-DRIVE, all negotiations are through the inter-vehicle negotiation, and the negotiation is always reliable and safe. The order of passing vehicles in the intersection correctly reflects the localization state and yielding cost.

\subsection{Perception-based A-DRIVE: For Mixed Traffic}

In this section, we present Perception-based A-DRIVE for the mixed traffic of CAVs and non-connected vehicles.
We firstly present a non-communication-based negotiation scheme that reflect the yielding cost.
In addition, we present the compatibility of V2V communication-based A-DRIVE and Perception-based A-DRIVE.

Since the negotiation for deadlock recovery cannot use the vehicular communications, each CAV shows its yielding cost by using its wait time.
The vehicle having the smaller cost starts to yield and recede first, and therefore, the vehicle having the larger cost may be consequently able to access the intersection area and resolve the deadlock.
To guarantee road safety around the intersection, each vehicle always uses its own perception systems and cannot proceed to avoid vehicle collisions.

\subsubsection{Non-communication-based Negotiation}\label{sec:non-comm_nego}

We now present a non-communication-based negotiation mechanism, in which each automated vehicle determines its threshold waiting time $\Delta$ by its yielding cost $\chi(v)$ presented in Eq. (\ref{eq:cost_func_percp}).
The threshold waiting time $\Delta_i$ for vehicle $i$ is the upper limit time period to wait in a deadlock situation. Once its waiting time is exceeding the threshold $\Delta_i$, the vehicle $i$ starts the backing operation.
The threshold time $\Delta_i$ is shown as follows.

\begin{equation}
\Delta_i = a \cdot \chi(v_i) + R_i
\label{eq:delta_function}
\end{equation}

Here, $a$ is a scale factor and $R_i$ is a random value generated at each vehicle. The random value $R_i$ is used by vehicle $i$ for breaking a tie when multiple vehicles involved in a deadlock situation have the exact same yielding cost.
This mechanism with random value for breaking a tie is also used for previous vehicle cooperation protocols \cite{azimi2013reliable, azimi2014stip}.

In this negotiation protocol, the vehicle being difficult to yield waits for long duration.
In other words, the vehicle being difficult to yield remains at the intersection and consequently accesses the intersection area first.
The vehicles being able to yield and recede easily have to yield immediately.
This logic is very straightforward to us, human drivers, and the protocol might be used even in a mixed traffic environment.

% This behavior is very straightforward to human drivers and the protocol can be used even in a mixed traffic, where human-operated vehicles and automated vehicles co-exist.

\subsubsection{Compatibility for 2 Components}

A-DRIVE includes $2$ components and they can be inclusively used for each CAV in practice.
In fact, the decision-making processes and the cost functions defined in Eqs. (\ref{eq:cost_func_comm}) and (\ref{eq:cost_func_percp}) have the multiple common elements.
In these $2$ subcomponents, each CAV calculates its threshold waiting time in a deadlock situation, and once its once its waiting time exceeds the given threshold, the vehicle starts the backing operation to yield the intersection. Consequently, the higher-priority vehicle can use the intersection area first.

In addition, A-DRIVE is robust and resilient against packet loss and/or delay, and never compromises road safety.
For example, when one of the vehicles involved in a deadlock situation has the communication failures and/or continuous packet losses, the vehicle is classified as a non-connected vehicle and the local perception-based protocol is used to calculate the threshold waiting time.
Since all connected vehicles cooperatively sense the presence of non-connected vehicle, all the vehicles consistently use the Local perception-based A-DRIVE in this case and no vehicle collisions and/or accidents occur.

%% file: 04_Implementation.tex
In this section, we present the implementation and an evaluation of A-DRIVE in the \textit{AutoSim}, a simulator-emulator \cite{bhat2018tools}.
We evaluate the deadlock recovery protocol in terms of traffic throughput, and compare to a simple baseline protocol.

\begin{figure}[!b]
\centering
\includegraphics[width=5.0cm]{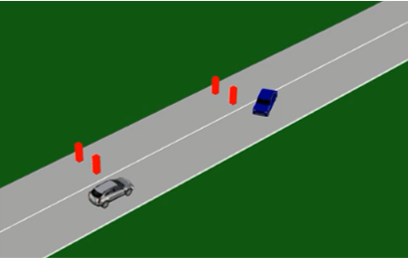}
\caption{Vehicle Deadlock in \textit{AutoSim}.}
\label{fig:VehicleDeadlock_AutoSim}
\end{figure}

AutoSim \cite{bhat2018tools} consists of various individual models, such as mobility, communication, control, localization, and pose estimation for each simulated vehicle.
In the simulator-emulator, each simulated CAV can transmit and receive a Basic Safety Message (BSM) and an Advanced Safety Message. The message format used is defined by the SAE J2735 standard \cite{J2735}, DSRC Message Set Dictionary.
AutoSim supports a realistic communication model and a packet loss model \cite{azimi2013reliable}, and we can evaluate our protocol with the realistic models.
Also, when vehicles conduct a backing operation, they simply go back the way they came, by using the map database and defined path.

We show a simple deadlock situation in AutoSim in Figure \ref{fig:VehicleDeadlock_AutoSim}. In this example scenario, the road segment has two lanes and one of them is closed due to the construction.
Two facing vehicles enter the construction zone at almost same time, and consequently lead the vehicle deadlock.
To recover from the deadlock, one of them has to recede and get out from the shared road segment.

\subsection{Performance Metric}

We define the \textit{Trip Time} that represents the time taken by a vehicle to traverse the intersection.
We calculate the trip time for each simulated vehicle and compare that against the trip time taken by the car assuming that it stays at a constant speed and does not stop at the intersection.
Here, trip time is defined as the time taken by a vehicle to go from a certain start-point before the shared road segment to a certain end-point after the shared road segment.
The difference between these two trip times is considered to be the \textit{Trip Delay} due to the intersection, and the \textit{Average Trip Delay} and \textit{Worst Trip Delay} across all CAVs in each simulation are compared in our evaluation.

\subsection{Scenarios}

\begin{figure}[!b]
  \begin{center}
    \begin{tabular}{c}
      % 1
      \begin{minipage}{0.5\hsize}
        \begin{center}
          \includegraphics[clip, width=4.35cm]{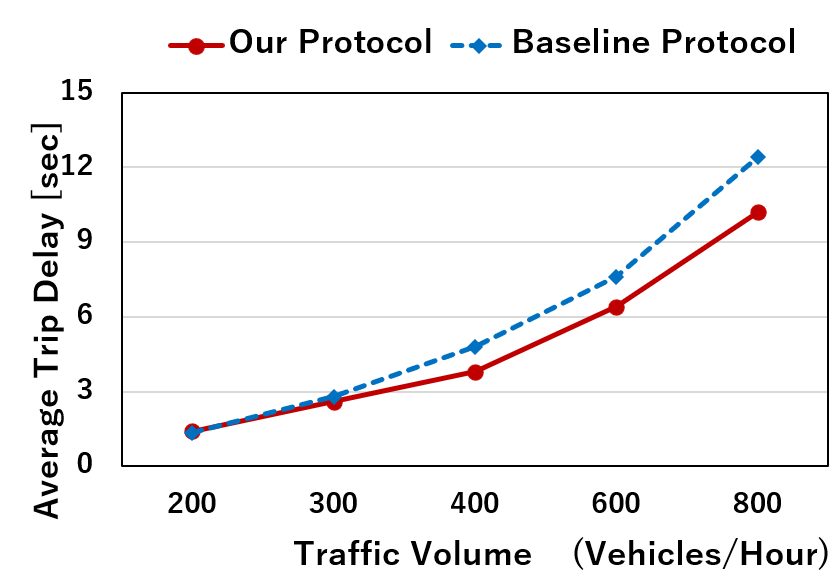}
	\hspace{1.6cm} (a) Average Trip Delay for Deadlock Recovery.
        \label{fig:case1}
        \end{center}
      \end{minipage}
      % 2
      \begin{minipage}{0.5\hsize}
        \begin{center}
          \includegraphics[clip, width=4.35cm]{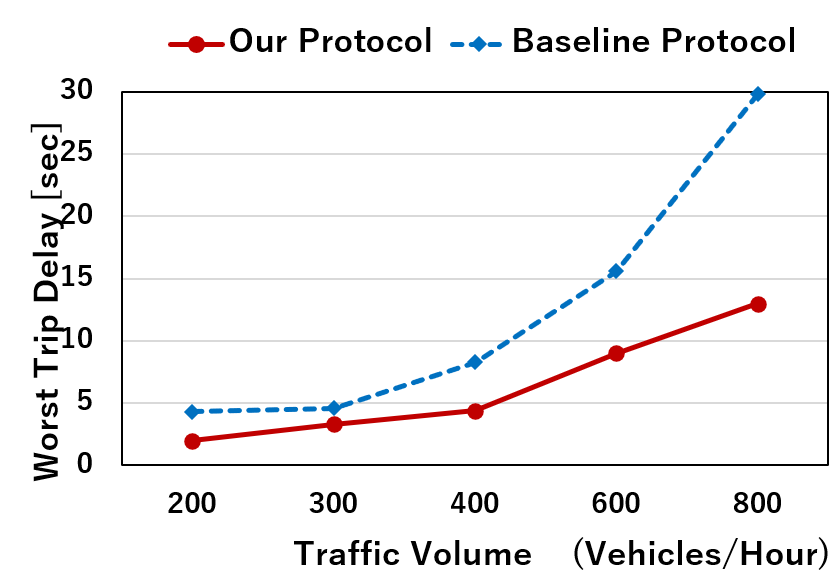}
	\hspace{1.6cm} (b) Worst Trip Delay for Deadlock Recovery.
          \label{fig:Deadlock_Worst}
        \end{center}
      \end{minipage}\\
      \begin{minipage}{0.10\hsize}
      \end{minipage}\\
    \end{tabular}
    \caption{Trip Delay and Traffic Volume.}
\label{fig:EvaluationOne}
  \end{center}
\end{figure}

We present the evaluation in a single-track lane, as shown in Figures \ref{fig:EvaluationOne} and \ref{fig:Delay_IntersectionSize}.
We show the single-track lane for the experiments in Figure \ref{fig:VehicleDeadlock_AutoSim}.
The variables for the simulations are set by utilizing the DSRC standards \cite{kenney2011dedicated, cheng2007mobile}. Also, the communication range and frequency for V2V communications are set as 400 m and 10 Hz, respectively.
Each simulation runs for 30 minutes, and we evaluate the time delay of the vehicles that are generated during the last 20 minutes to skip initial transients and measure steady-state behavior.
Our simulations evaluate A-DRIVE by comparing against one baseline traffic protocol: \textit{Lane Priority-based Protocol}.

% In addition, in this evaluation, we do not use vehicle cooperation protocols to evaluate the effect of the deadlock recovery protocol. Each vehicle greedily enters the single-track lane, and the vehicle uses our A-DRIVE once it is involved in the deadlock.

%\vspace{1.2mm}
\begin{description}
 \item[\textbf{Lane Priority-based Protocol}]\mbox{}\\ 
Vehicles follow the lane priority that is given from the map database.
Even when one vehicle stops near the intersection exit, the vehicle has to start the backing operation if it drives on the lower-priority lane.
\end{description}

To evaluate our protocol from sparse traffic rural road segments to very congested urban road segments, we change the traffic volume from 200 vehicles per hour to 800 vehicles per hour for each direction.
Also, in the simulation, the traffic arrival pattern follows the exponential distribution.
In addition, we change the size of intersection from 10 meters to 100 meters.

\subsection{Evaluation for Traffic Throughput}

\begin{figure}[!t]
\centering
\includegraphics[width=7.5cm]{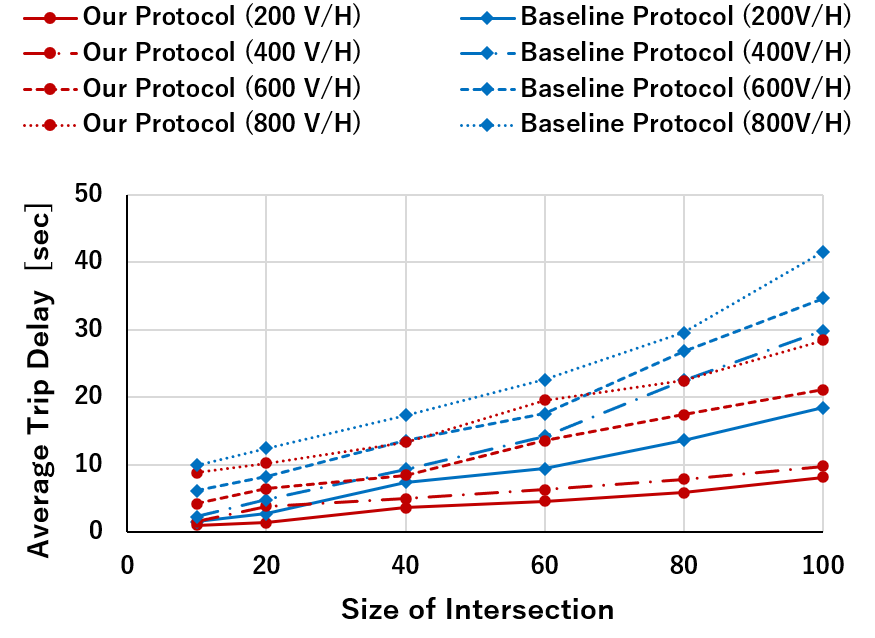}
\caption{Trip Delay and Intersection Size.}
\label{fig:Delay_IntersectionSize}
\end{figure}

We first evaluate the Average Trip Delay and the Worst-case Trip Delay by changing the traffic volume, as shown in Figure \ref{fig:EvaluationOne}.
Here, in all cases, A-DRIVE demonstrates the superior performance, compared to the baseline protocol. When the packet losses happen, the vehicles start to use the Local perception-based A-DRIVE.
When the traffic volume is low, A-DRIVE and the simple baseline protocol demonstrate similar performance. On the other hand, when the traffic volume increases, the lane priority-based protocol has longer worst-case waiting time, because the vehicles on lower-priority lane might have to keep yielding and waiting for a long time.

In addition, we show the Average Trip Delay by changing the physical size of intersection and the traffic volume, as shown in Figure \ref{fig:Delay_IntersectionSize}.
In all cases, A-DRIVE demonstrates the better performance than the baseline protocol.
In the baseline protocol, the vehicles on lower-priority lane sometimes have to recede multiple times.
In particular, for the long intersection, the recovery times from deadlocks become long time and the trip delays are exponentially increased.
On the other hand, in A-DRIVE, each CAV negotiates to determine the priority and does not need to take these unnecessary actions.

%% file: 05_relatedwork.tex
Deadlock properties have widely studied in computer systems \cite{holt1972some}, distributed systems \cite{singhal1989deadlock}, and database systems \cite{menasce1979locking} to detect and resolve deadlocks to improve the systems performance.
Many researchers extended these studies to collision- and/or deadlock-free coordination for multi-robot systems \cite{jager2001decentralized, li2005motion} and intelligent vehicle systems \cite{aoki2018dynamic, liu2018distributed}, in order to improve safety and/or enhance traffic throughput.
These works used different approaches, such as a resource management, a priority assignment, and a multi-robot motion planning, but one common feature is focusing on the shared road segment that may lead collisions and deadlocks.

In particular, road intersections are serious bottlenecks for transportation and many researchers focused on collision- and/or deadlock-free coordination at intersections.
First, Dresner and Stone proposed a multi-agent approach for an intersection control, named Autonomous Intersection Management \cite{dresner2005multiagent}.
In the system, all vehicles call ahead to a reservation manager agent at the intersection to reserve conflict-free trajectories. 
Khayatian et al. \cite{khayatianrim2018RIM} extended the system to robust against external disturbances and model mismatches.
Also, Azimi et al. \cite{azimi2013reliable, azimi2014stip} proposed several spatio-temporal protocols for road intersections.
In this work, vehicles are assigned priorities based on their arrival times at the intersection, with vehicles reaching the intersection earlier assigned higher priorities than vehicles coming later.
In addition, a novel family of synchronous intersection protocols, BRIP \cite{azimi2015ballroom}, CSIP \cite{aoki2017configurable}, and DSIP \cite{aoki2019DSIP} were presented.
Under these synchronous intersection protocols, all automated vehicles synchronously cross the intersection without stopping by following a strict but well defined spatio-temporal pattern tailored to each intersection.
Lin et al. \cite{lin2019graph} derived formal verification approaches based on graphs and Petri nets for deadlock-free intersection management.

Many other driving contexts, such as single-track lanes, merge points, lane-changing, passing, and construction zones \cite{aoki2017merging, zhao2019state, xiao2019decentralized, liu2018dynamic, aoki2018dynamic, hakeem2017cooperative, aoki2021multicruise}, often require collision- and deadlock-free cooperation and collaboration.
In \cite{aoki2018dynamic}, these driving contexts were defined as Dynamic Intersections, and the cooperative protocol for dynamic intersections and Cyber Traffic Light were proposed.
In the protocol, each CAV negotiates the priority for the shared road segment by using V2V communications and dynamically determine the priority based on the surrounding situations, including traffic volume and waiting time.
In addition, Liu et al. \cite{liu2018dynamic} decomposed the lane-changing maneuver to two parts: a longitudinal control problem and an associated lateral control.
They also partitioned the road segment into a longitudinal segment, a lateral segment, and a lane-keeping segment, and developed vehicle movement models for each segment.
These studies presented decision-making mechanisms for many different types of driving contexts, but did not consider how multiple vehicles recover from the vehicle deadlock.
Unlike these previous works, our study considers the vehicle system includes reasonable failures and the vehicle deadlocks might happen almost anywhere and anytime on public roads.

%% file: 06_conclusion.tex
In this paper, we presented a practical and cooperative deadlock recovery protocol named \textbf{A-DRIVE} for Connected and Autonomous Vehicles (CAVs) to safely recover from the deadlocks at a stationary/dynamic intersection.
In A-DRIVE, to co-exist with non-connected automated vehicles and human-driven vehicles, each CAV uses both V2V communication-based A-DRIVE and Local perception-based A-DRIVE in parallel.
These $2$ components use the common cost function and same logic for the decision-making processes, and CAVs can complementarily and inclusively use these components.
Hence, CAVs do not need to change their behaviors by the presence of human-driven vehicles, and resolve the deadlock situations.
In the evaluation, we showed that our protocol shortens the recovery time from the vehicle deadlocks and increases the traffic throughput at a road intersection, a merge point, a single-track lane, and a construction zone.

We finally note several limitations of our work.
First, to reduce the number of vehicle deadlocks on public roads, we have to consider cooperative navigation and route planning. If we can coordinate all the self-driving vehicles at the navigation level, the vehicles can reduce the chances of even encountering each other.
In addition, we assume that a long single-track lane has reasonable traffic volumes and that the evacuation spaces are not the bottleneck.
In future work, we will consider the priority assignment that accounts for the space constraints of passing places and evacuation spaces.